\documentclass[twocolumn,showpacs,amssymb,aps,prc,nofootinbib,floatfix,superscriptaddress]{revtex4-1}

\usepackage{supertabular}
\usepackage{lipsum}
\usepackage{blindtext}
\usepackage{amsfonts}
\usepackage{tensor}
\usepackage{graphicx}
\usepackage{pict2e}
\usepackage{balance}
\usepackage{blindtext}

\usepackage{amssymb, amsmath,environ}
\usepackage[normalem]{ulem}

\usepackage{color}
\usepackage[colorlinks=true,urlcolor=black,linkcolor=black,citecolor=red]{hyperref}


\newcommand{\bse}{\begin{subequations}}
	\newcommand{\ese}{\end{subequations}}
\newcommand{\be}{\begin{equation}}
\newcommand{\ee}{\end{equation}}

\NewEnviron{eqsplit}{%
	\begin{equation}\begin{split}
	\BODY
	\end{split}\end{equation}
}

\makeatletter
\newcommand*\bigcdot{\mathpalette\bigcdot@{.5}}
\newcommand*\bigcdot@[2]{\mathbin{\vcenter{\hbox{\scalebox{#2}{$\m@th#1\bullet$}}}}}
\makeatother

\newcommand{\bea}{\begin{eqnarray}}
\newcommand{\eea}{\end{eqnarray}}
\newcommand{\ba}{\begin{array}}
	\newcommand{\ea}{\end{array}}

\newcommand{\la}{\langle}
\newcommand{\ra}{\rangle}

\begin{document}

\title{Charm balance function in relativistic heavy-ion collisions}

\author{Tribhuban Parida}
\email[]{tribhubanp18@iiserbpr.ac.in}
\affiliation{Department of Physical Sciences, Indian Institute of Science Education and Research Berhampur, 
Transit Campus (Govt ITI), Berhampur-760010, Odisha, India}

\author{Piotr Bo\.zek}
\email[]{Piotr.Bozek@fis.agh.edu.pl}
\affiliation{AGH University of Krakow, Faculty of Physics and Applied Computer Science, 
aleja Mickiewicza 30, 30-059 Cracow, Poland}

\author{Sandeep Chatterjee}
\email[]{sandeep@iiserbpr.ac.in}
\affiliation{Department of Physical Sciences, Indian Institute of Science Education and Research Berhampur, 
Transit Campus (Govt ITI), Berhampur-760010, Odisha, India}

\begin{abstract}
We calculate the balance function for charm in relativistic heavy-ion collisions. The distribution of pairs of 
charm-anticharm quarks produced in hard processes in the early stages of the nucleus-nucleus collision evolves 
in the dense fireball formed in the collision. The evolution of the dense matter is described using a relativistic viscous 
hydrodynamic model and the quark diffusion with a Langevin equation. The evolution of the charm quark balance 
function from the formation of the charm-anticharm pair up to the freeze-out traces the partial thermalization of 
the heavy quarks in the dense matter. For the balance function in azimuthal angle we reproduce the  collimation effect 
due to the transverse flow. The evolution in rapidity shows the thermalization of the longitudinal velocity of the quark in 
the fluid.  We provide  estimates for the one and two-dimensional balance functions for $D^0$-$\bar{D^0}$ mesons produced 
in ultarelativistic Pb+Pb collisions at $\sqrt{s_{NN}}=5.02$ TeV. The shape of the charm balance function in relative 
rapidity is sensitive to the rescattering of heavy quarks in the early stages of the collision, while the shape of the 
balance function in azimuthal angle is sensitive to the rescattering in the latter stages.
\end{abstract}


\maketitle

\section{Introduction}

Heavy quarks are produced in hard processes at the beginning of relativistic nuclear collisions. The interaction of such 
particles produced at the early stages with the surrounding medium makes them an excellent probe of the dynamics of 
the dense matter created in the interaction region of the collision \cite{Svetitsky:1987gq,Prino:2016cni,Dong:2019byy,
Dong:2019unq,He:2022ywp}. Measurements of the nuclear modification factor and the harmonic flow of open charm mesons 
provide an insight into the partial thermalization of hard produced charm quarks.

The charge balance function has been proposed as a measure of the interaction of charged quarks in the  quark-gluon 
plasma \cite{Bass:2000az}. The two opposite charges produced in the plasma would drift apart in relative rapidity due 
to rescattering with the thermal medium. The charge balance function is defined in a way to identify the specific correlated 
charge-anticharge pairs. The balance function in the relative azimuthal angle is sensitive to the amount of transverse flow 
at the time of the charge pair separation \cite{Bozek:2004dt}. The balance function involving the electric charge, strangeness 
and baryon number can be measured \cite{STAR:2003kbb,STAR:2010plm,NA49:2004vzs,NA49:2007mss,ALICE:2013vrb,
CMS:2023sua}. Charges carried by light quarks and hadrons can be produced during the entire span of the collision. In 
particular, a large part of the balance function for the electric charge receives contribution from local charge conservation 
for charges produced at hadronization \cite{Jeon:2001ue}. The full description of the charge balancing dynamics combines 
the charge-anticharge pairs produced in the medium and at the hadronization \cite{Pratt:2011bc,Pratt:2012dz,Pratt:2018ebf}. 
Charm quarks are almost exclusively produced in the initial hard scatterings. The two balancing charm-anticharm quarks 
evolve in the medium. It has been noted that the charm-anticharm correlations or the charm  balance function could serve 
as a clean probe of such interactions \cite{Zhu:2006er,Basu:2021dzv}. For example, the charm balance function could be sensitive to the formation of charmonium bound states in the quark-gluon plasma.

In this paper, we study the charm-anticharm balance function in relative azimuthal angle and relative rapidity. The heavy quarks 
produced in hard processes  evolve in the quark-gluon plasma, which changes their relative momenta. In comparison to the 
dynamics of light quarks, the hard produced heavy quarks have large initial momenta. In our calculation, we take charm and 
anticharm quarks as produced in p+p collisions using the Pythia model \cite{Bierlich:2022pfrpub}. The quarks rescatter in the dense 
matter of the fireball. In relative azimuthal angle the balance function becomes narrower. This is equivalent to the effect of the 
increase of the correlation function of the final  $D^0$ and $\bar{D}^0$ mesons at small relative azimuthal angle
 \cite{Zhu:2006er,Zhu:2007ne,Younus:2013be,Tsiledakis:2009qh,Nahrgang:2013saa,Adare:2013wca}. The balance function in 
 relative rapidity becomes broader after the diffusion of the heavy quarks in the medium. The behavior of the  charm balance 
 in relative rapidity provides an additional information, because it is sensitive to the early stages of the reaction, unlike the balance 
 function in azimuthal angle. 

In the next section we describe the calculation of the balance function in the model. Then, in Sec. \ref{sec:cqbf} the balance function 
is calculated directly for the charm quarks before and after rescattering in the medium. In Sec. \ref{sec:ccevolution} we study the 
time evolution of different contributions  leading to the broadening of the charm balance function in rapidity. Calculations of  the 
balance function for $D^0$-$\bar{D}^0$ mesons are provided in Sec. \ref{sec:DD}. A simple estimate of possible effects of a 
pre-hydrodynamic phase on the charm balance function is given in Sec. \ref{sec:early}. Conclusions about the charge balance 
functions in relative azimuthal angle and relative rapidity are summarized in the last section.

\section{Calculation of the charm balance function in the hydrodynamic model}

\begin{figure}[th!]
	\includegraphics[scale=0.7]{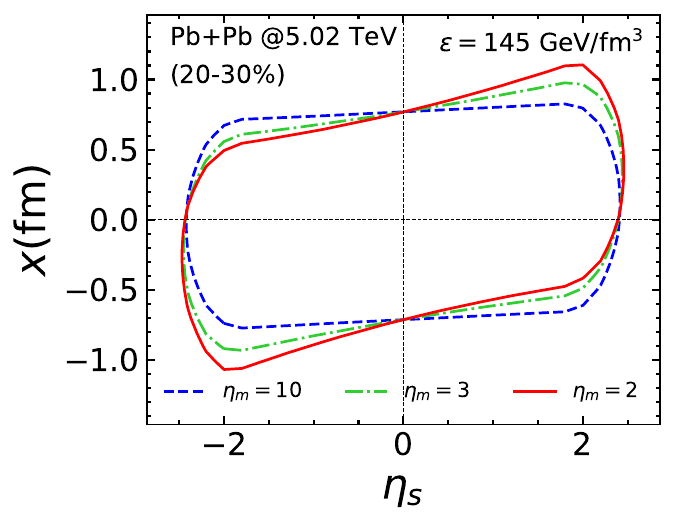} 
  \caption{ Contour plots depicting constant energy density at $\tau = \tau_0$ have been plotted for various tilt parameters ($\eta_m$) in Pb+Pb 
collisions at a center-of-mass energy of $\sqrt{s_{NN}} = 5.02$ TeV and an impact parameter of $b = 7.7$ fm. The contours are plotted in the 
$x-\eta_s$ plane at $y=0$. }
	\label{fig:tilt_profile}
\end{figure}

We have adopted the Optical Glauber model to set up the initial condition for the hydrodynamic evolution. The initial three 
dimensional distribution of the energy density ($\epsilon$) at a constant proper time ($\tau_{0}$) takes the following form:
\begin{eqnarray}
\epsilon(x,y,\eta_{s}) &=& \epsilon_{0} \left[ \left( N_{+}(x,y) f_{+}(\eta_{s}) + N_{-}(x,y) f_{-}(\eta_{s})  \right)\right. \nonumber\\
                           &&\left.\times \left( 1- \alpha \right) + N_{coll} (x,y)  \epsilon_{\eta_s}\left(\eta_{s}\right) \alpha \right]
\label{eq_tilt}
\end{eqnarray}
where, $N_{+}(x,y)$ and $N_{-}(x,y)$ are the participant densities from the forward and backward going nuclei. $N_{coll}(x,y)$ is 
the density of the binary collision sources evaluated at any transverse position (x, y). $\alpha$ is the hardness factor and 
$\epsilon_{0}$ is the scaling parameter of energy density. $\epsilon_{\eta_s}\left(\eta_{s}\right)$ and $f_{+,-}(\eta_s)$ are the 
profiles that determine the deposition of $\epsilon$ in space-time rapidity ($\eta_s$).  The forms of 
$ \epsilon_{\eta_s}\left(\eta_{s}\right)$ and $f_{+,-}(\eta_s)$ are given in Eq.~\ref{eq_etas_even_profile_for_epsilon} and 
Eq.~\ref{eq_etas_odd_profile_for_epsilon} respectively,
\begin{equation}
  \epsilon_{\eta_s}(\eta_s) = \exp \left(  -\frac{ \left( \vert \eta_{s} \vert - \eta_{0} \right)^2}{2 \sigma_{\eta}^2}   
    \theta (\vert \eta_{s} \vert - \eta_{0} ) \right) \ ,
    \label{eq_etas_even_profile_for_epsilon}
\end{equation}

\begin{equation}
    f_{+,-}(\eta_s) = \epsilon_{\eta_s}(\eta_s) \epsilon_{F,B}(\eta_s) \ ,
    \label{eq_etas_odd_profile_for_epsilon}
\end{equation}
where
\begin{equation}
    \epsilon_{F}(\eta_s) = 
    \begin{cases}
    0, & \text{if } \eta_{s} < -\eta_{m}\\
    \frac{\eta_{s} + \eta_{m }}{2 \eta_{m}},  & \text{if }  -\eta_{m} \le \eta_{s} \le \eta_{m} \\
    1,& \text{if }  \eta_{m} < \eta_{s}
\end{cases}
    \label{eq_etas_odd_profile_for_epsilon_2}
\end{equation}
and 
\begin{equation}
    \epsilon_{B} (\eta_s) = \epsilon_F(-\eta_s) \ .
\end{equation}
$f_{+,-}(\eta_s)$ introduces an asymmetric deposition of energy density by a participant source along $\eta_s$. This kind of 
initial condition generates a tilted profile in the reaction plane (plane made by beam axis and impact parameter) of the 
collision~\cite{Bozek:2010bi}. The strength of the initial tilt in the model can be controlled by the parameter $\eta_m$ present 
in Eq.~\ref{eq_etas_odd_profile_for_epsilon_2}.
\textcolor{blue}{
In order to illustrate the tilted profile and elucidate the effect of the tilt parameter $\eta_m$ on the initial energy distribution, we have generated contour plots representing constant energy density ($\epsilon = 145$ GeV/fm$^3$) in the $x-\eta_s$ plane at $y=0$ for different values of $\eta_m$. These results are presented in Fig.~\ref{fig:tilt_profile}.
}
We have used $\alpha = 0.14 , \eta_{0}=1.8 $, $\sigma_{\eta}=2.0$ and 
$\eta_m=3.2$ that describe the experimental data of the charged particle yield and directed flow in Pb+Pb collisions at 5.02 TeV.

The viscous hydrodynamic evolution of the energy density profile has been performed by the publicly available MUSIC code 
\cite{Schenke:2010nt,Schenke:2010rr,Paquet:2015lta}. We have taken shear viscosity to entropy density ratio 
$\eta/s=0.04$~\cite{Alba:2017hhe, Giacalone:2017dud} and have used a lattice QCD based Equation of State~\cite{Monnai:2019hkn} 
during the hydrodynamic evolution.

  In this framework, we conduct a single-shot hydrodynamic evolution of the initial energy density obtained from the optical Glauber model calculation for a specific centrality class. The initial energy density is computed with a specific impact parameter $b$ value. 
  We employ the average impact parameter value from the $b$ distribution in the Glauber model calculation to determine the initial energy density distribution for each centrality. For the centrality classes of 10-20\%, 20-30\%, 30-40\%, 40-50\%, 50-60\%, and 60-70\%, the corresponding impact parameter values used are 6.0, 7.7, 9.2, 10.4, 11.5, and 12.5 fm, respectively.

In our simulation of a specific centrality class for Pb+Pb collisions at a center-of-mass energy of $\sqrt{s_{NN}} = 5.02$ TeV, we initiate the process with the Pythia~\cite{Bierlich:2022pfrpub} simulation of p+p collisions.  
We take all the charm-anticharm pairs generated from $5\times 10^{6}$ Pythia events of p+p collisions at a center-of-mass energy of $\sqrt{s_{NN}} = 5.02$ TeV. Subsequently, we initialize the positions of the produced charm and anticharm quarks within the fluid modeled by hydrodynamics and execute independent Langevin evolutions of each quark. 
The initial momentum of the quarks are kept the same as obtained from Pythia.
Assuming that charm and anticharm quarks are produced through initial hard scatterings, we sample the initial transverse positions for each pair from binary collision profiles denoted as $N_{coll}(x,y)$. We set the initial space-time rapidity $\eta_s$ equal to the rapidity of the respective quarks. Consequently, the initial space-time rapidity difference $\Delta \eta_s$ between a pair of charm and anticharm quarks is identical to the initial rapidity difference $\Delta y$ between them.

The hydrodynamic evolution provides us the relevant dynamic properties of the background (local temperature $T$ and flow velocity 
$u^\mu$), in which the charm and anticharm quarks evolve. The quark evolution stops once the local background temperature falls below 
150 MeV. The dynamics of the heavy quarks in the medium is described using the  Langevin equation
\begin{eqnarray}
\Delta {\bf r}_i &=& \frac{{\bf p}_i}{E}\Delta t\label{eq.Langevinx} \nonumber \\
\Delta {\bf p}_i &=& -\gamma {\bf p}_i\Delta t + \rho_i\sqrt{2D\Delta t} \ ,
\label{eq:Langevin}
\end{eqnarray}
where $i \in \{ x,y,z \} $ refers to the  three  Cartesian coordinates.
$\Delta {\bf r_i}$ and $\Delta {\bf p_i}$ are the   updates of the coordinates in position and momentum of the heavy quark 
respectively within the  time interval  $\Delta t$.  $\gamma$ and $D$ are the drag and diffusion coefficients, which effectively 
describe  the interactions between the heavy quark and the medium. We take $D=\gamma E T$, where $E=\sqrt{\vert p \vert ^2+m_c^2}$ is the energy of the heavy quark with mass $m_c$. We  assume a diagonal form for the diffusion matrix~
\cite{vanHees:2005wb,Cao:2011et,Scardina:2017ipo}. The noise term $\rho_i$ is randomly sampled from a normal distribution at 
every time step, with $\la\rho_i\ra=0$ and $\la\rho_i\rho_j\ra=\delta_{ij}$.
We have followed the post point scheme prescribed in Ref.~\cite{He:2013zua} to update the momentum in the Langevin evolution. This scheme ensures 
that the heavy quark phase space distribution approaches to the equilibrium Boltzmann-Juttner distribution after long time evolution.
We consider 
$\gamma = 0.4 T$, that allows to capture the suppression factor  and the elliptic flow of heavy quarks \cite{Chatterjee:2017ahy}.

The balance function is defined as a conditional probability isolating  correlations between balancing charges 
\cite{Bass:2000az,Jeon:2001ue,Schlichting:2010qia}
\begin{eqnarray}
  B(p_1|p_2)&=&\frac{1}{2}\left[\frac{N^{c\bar{c}}(p_1,p_2)}{N^{\bar{c}}(p_2)}-\frac{N^{\bar{c}\bar{c}}(p_1,p_2)}{N^{\bar{c}}(p_2)} \right. \nonumber \\
    && \left. + \frac{N^{\bar{c}c}(p_1,p_2)}{N^{{c}}(p_2)}-\frac{N^{{c}{c}}(p_1,p_2)}{N^{{c}}(p_2)} \right] \ ,
\end{eqnarray}
where  $N^a(p_1)$ and $N^{ab}(p_1,p_2)$ are the one and the two-particle distributions of particles of type $a, b$ with 
momenta $p_1$ and $p_2$. The balance functions have been studied in models and in experiments for the electric charge, 
strangeness and baryon number \cite{STAR:2003kbb,STAR:2010plm,NA49:2004vzs,NA49:2007mss,ALICE:2013vrb,CMS:2023sua}. 
Theses charges can be produced  in the initial nucleon-nucleon scattering,  in the evolving  hot and dense fireball, and during the 
hadronization process. The understanding of  the charge balancing in heavy-ion collisions requires in that case a careful modeling 
of the different ways of the charge-anticharge production \cite{Pratt:2011bc,Pratt:2012dz,Pratt:2015jsa,Pratt:2021xvg,Pruneau:2023zhl,
Pruneau:2022brh,Pratt:2022kvz}. However, owing to their large mass, the charm quarks are expected to be produced only during the 
initial hard scatterings.

Assuming the independent evolution of $c$-$\bar{c}$ pairs created in different hard process, one can write balance function using 
the correlation function calculated for each $c$-$\bar{c}$ pair separately. In this work, we do not take into account possible correlations due to the formation and dissolution  of charmonium bound states \cite{Basu:2021dzv}. The quark and antiquark  are evolved in the bulk matter 
according to Langevin equation. The two-particle distribution $\rho^{c\bar{c}}(p_1,p_2) $ and the one-particle distributions $\rho^c(p_1)$, 
$\rho^{\bar{c}}(p_1)$ of hadrons containing the charm quark and antiquark can be obtained as an average over all the pairs. These 
distributions are normalized to one:
\begin{equation}
  \int dp_1 dp_2 \rho(p_1,p_2) = \int dp_1 \rho^c(p_1)= \int dp_1 \rho^{\bar{c}}(p_1)=1 \ .
  \end{equation}
    The two-dimensional balance function can be easily constructed as
    \begin{eqnarray}
      B(\Delta y,\Delta \phi) = \int_A dp_1 \int_A dp_2 & &  \nonumber  \\ \frac{\rho^{c\bar{c}}(p_1,p_2) \delta\left(\Delta y - y_1+
      y_2\right)\delta\left(\Delta \phi -|\phi_1-\phi_2|\right) }{\int_A dp_1\rho^c(p_1)}  & & \ ,
    \end{eqnarray}
    where we assume a symmetric distribution between $c$ and $\bar{c}$ quarks and use a symmetric choice of the acceptance region 
    $A$ in azimuthal angle and rapidity. In this paper we study the balance function in  relative azimuthal angle and/or in relative rapidity, 
    where the acceptance region is given by cuts on the transverse momentum of the particles. If the region $A$ involves all particles 
    containing $c$ and $\bar{c}$ quarks and all momenta, the balance function is normalized
    \begin{equation}
      \int_{-\infty}^\infty d (\Delta y) \int_0^\pi d(\Delta \phi) B(\Delta y,\Delta \phi)=1 \ .
      \label{eq:balnorm}
    \end{equation}
    With limited kinematic acceptance and efficiency or if not all open charm hadrons  and charmonia are accounted for, the normalization 
    is $<1$.
    
\section{Balance function for $c$-$\bar{c}$ quarks }
\label{sec:cqbf}

\begin{figure}[th!]
	\includegraphics[scale=0.4]{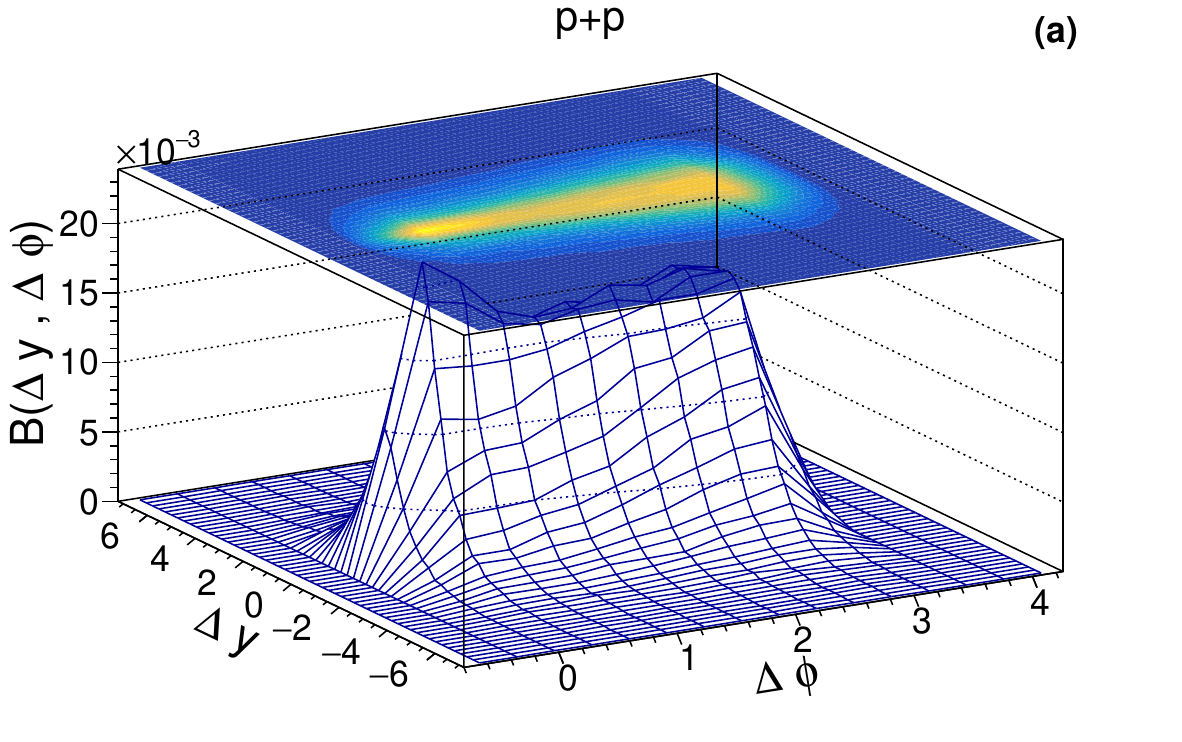} \\
	\includegraphics[scale=0.4]{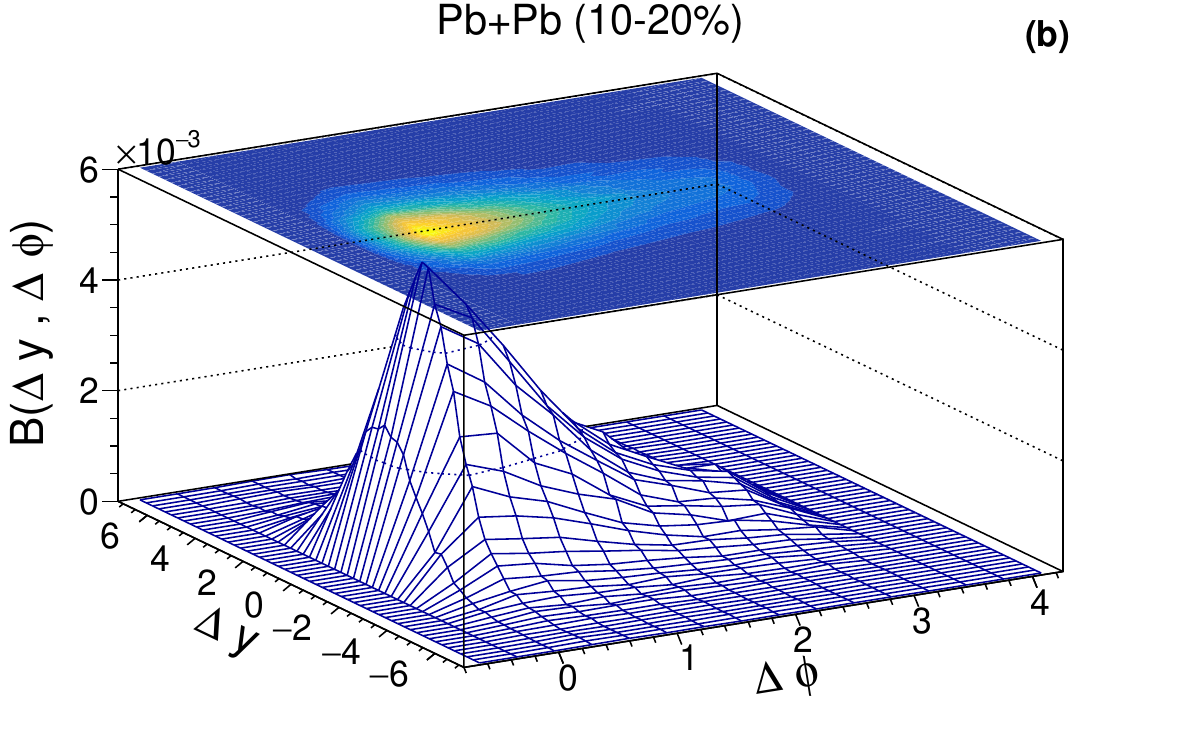} 
	\caption{The two dimensional balance function for $c$-$\bar{c}$ quarks in relative rapidity $\Delta y$ and relative azimuthal angle 
	$\Delta \phi$ for $p_T > $ 2 GeV. The panel (a) represents simulation results from Pythia for quarks emitted in p+p collisions and the 
	panel (b) shows the balance function after the evolution of the quark-antiquark pairs in the bulk matter formed in Pb+Pb collisions with $10-20\%$
	centrality. } 
	\label{fig:2Dbal}
\end{figure}

In this section we study the balance functions  for the $c$-$\bar{c}$ quarks to understand the evolution of the balance function in the 
collision. The two-dimensional balance function for quark pairs emitted from p+p collisions at $\sqrt{s_{NN}}=5.02$ TeV from the 
Pythia model is shown in  panel (a) of Fig. \ref{fig:2Dbal}. The correlation has a broad distribution in $\Delta \phi$ with a maximum 
for back-to-back pairs. In the panel (b), we show the balance function for $c$-$\bar{c}$ pairs emitted after Langevin diffusion 
in the dense fireball formed in Pb+Pb collisions with $10-20\%$ centrality.

After diffusion in the bulk matter the quark-antiquark momenta  pick up a contribution of the transverse flow, as a result the azimuthal 
distribution becomes more peaked for pairs flying in the same direction. The longitudinal distribution in rapidity is  broadened due to the rescattering 
in the medium.
  
  
\begin{figure}[th!]
	\includegraphics[scale=0.4]{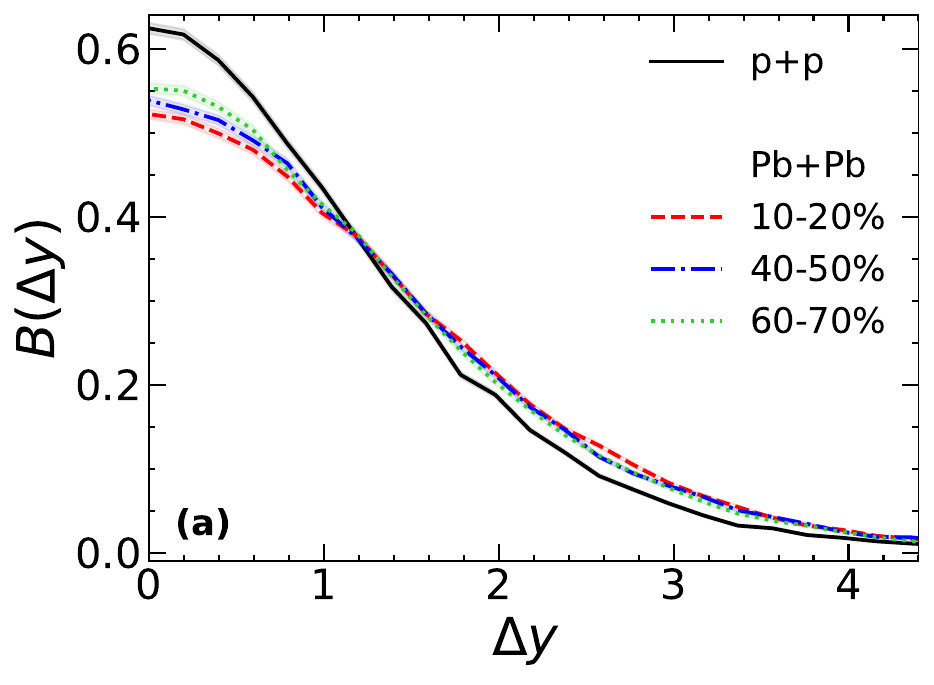} 
	\includegraphics[scale=0.4]{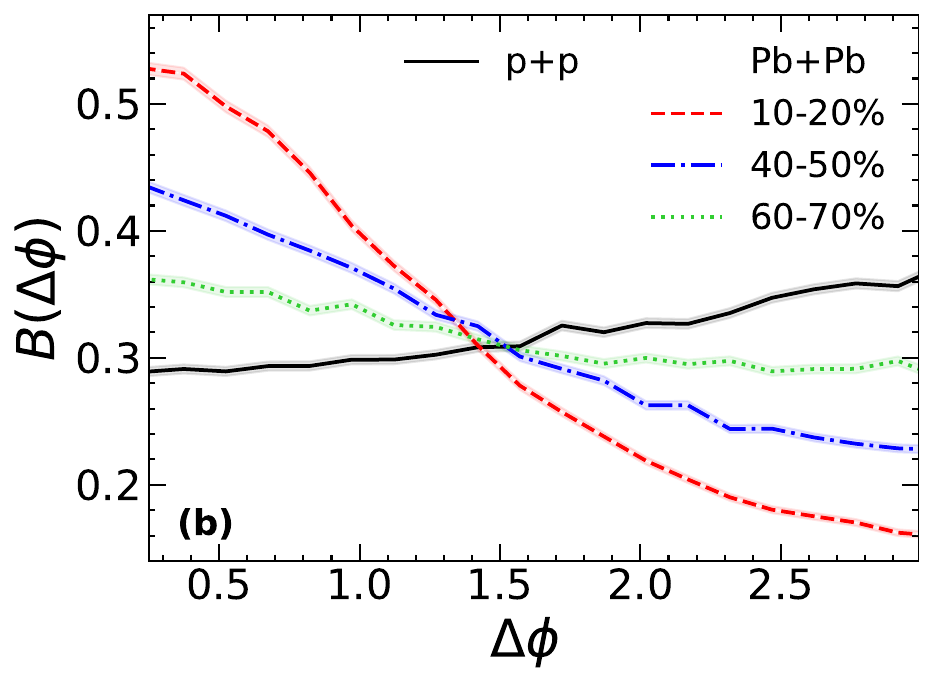}
	\caption{The balance function for $c-\bar{c}$ quarks in (a) relative rapidity and 
        in (b) relative azimuthal angle for p+p collisions (solid line) and 
	Pb+Pb collisions with centrality $10-20\%$ (dashed line), $40-50\%$ (dashed-dotted line), and $60-70\%$ (dotted line). Quarks with 
	any momentum are included in the balance function. } 
	\label{fig:ccbaly}
        \end{figure}


In the following we use the  one dimensional balance function in relative rapidity,

\begin{equation}
  B(\Delta y)=2 \int_0^\pi d(\Delta \phi) B(\Delta y, \Delta \phi) \ ,
\end{equation}

and relative azimuthal angle

\begin{equation}
  B(\Delta \phi)= \int_{-\infty}^\infty d(\Delta y) B(\Delta y, \Delta \phi) \ .
\end{equation}

Due to symmetry we define the one-dimensional balance functions $B(\Delta y)$ and $B(\Delta \phi)$  in the ranges
$y\ge 0$ and $[0$,$\pi]$ respectively.

The balance function $B(\Delta y)$ is shown in Fig. \ref{fig:ccbaly}(a). The plots
indicate that the balance function for quarks undergoing rescattering in the medium created in Pb+Pb collisions become 
broader as compared to the balance function for quarks emitted in  p+p collisions. There is some dependence of the shape 
of the balance function on collision centrality in Pb+Pb collisions. The balance function in relative azimuthal angle $B(\Delta \phi)$ 
shows a qualitative difference between p+p and Pb+Pb collisions (Fig. \ref{fig:ccbaly}(b)). The balance function is peaked at 
$\Delta \phi=\pi$ in p+p collisions (back-to-back pairs), while in Pb+Pb collisions it is peaked at $\Delta \phi=0$ (collimated pairs). 
The peak of the balance function at $\Delta \phi=0$ is an effect of the transverse flow. We recover the known collimation of 
the heavy quarks due to the rescattering in the expanding medium \cite{Zhu:2006er,Zhu:2007ne,Younus:2013be,Tsiledakis:2009qh,
Nahrgang:2013saa,Adare:2013wca}.

\begin{figure}[th!]
	\includegraphics[scale=0.4]{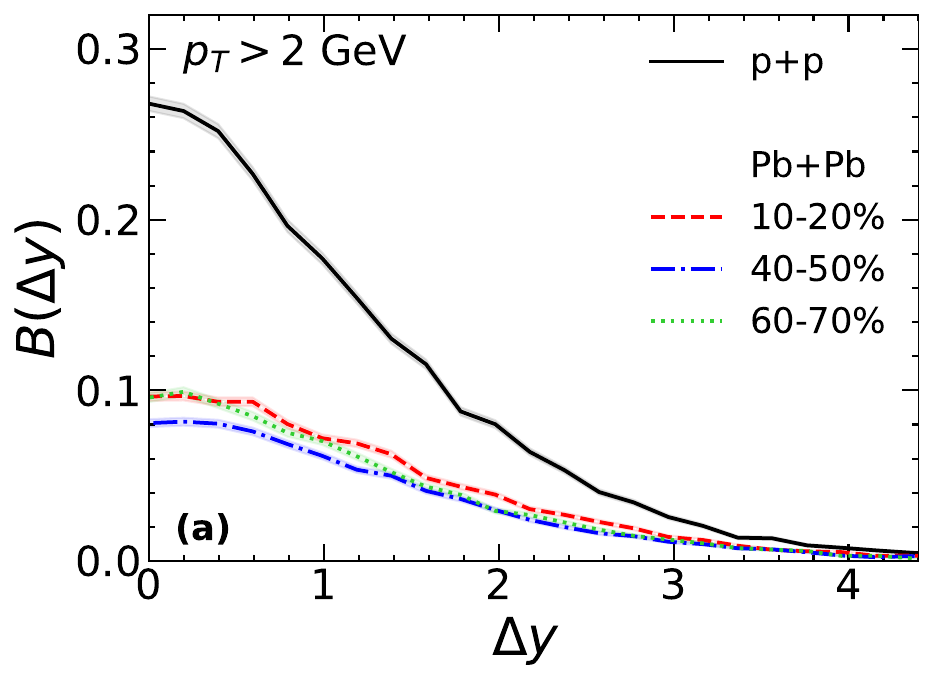} \\
	\includegraphics[scale=0.4]{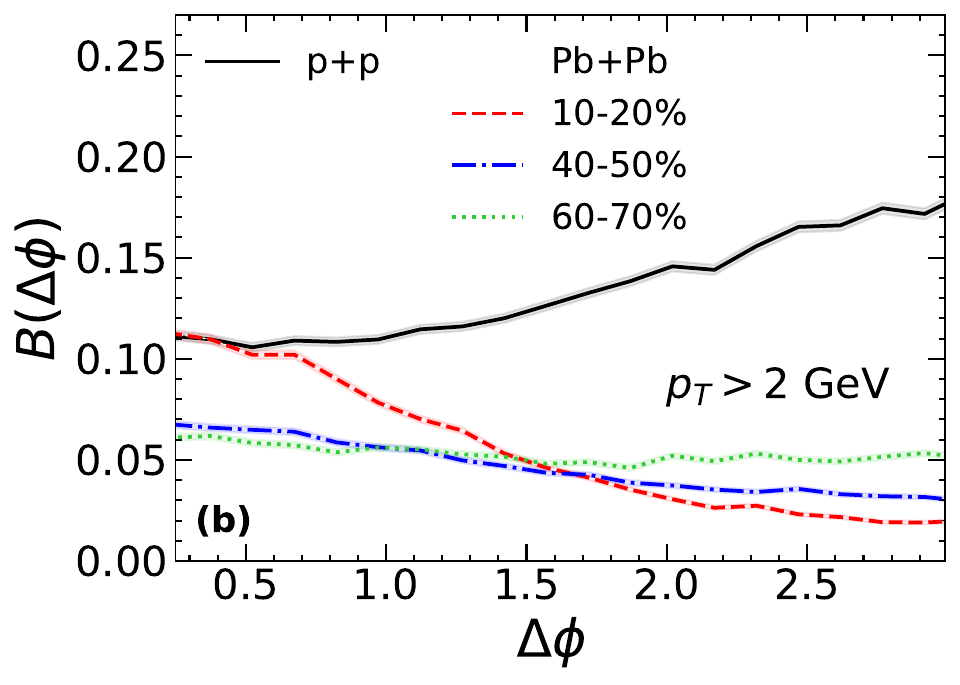}
	\caption{Same as in Fig. \ref{fig:ccbaly} but for quarks and antiquarks with $p_T>2$ GeV.}
	\label{fig:ccbaly2}
        \end{figure}

The one-dimensional balance functions for quarks and antiquarks with $p_T>2$ GeV are shown in Figs. \ref{fig:ccbaly2}(a) 
and \ref{fig:ccbaly2}(b). The  cut $p_T>2$ GeV makes the balance function in relative rapidity narrower. The effect of the 
$p_T$ cut on the  balance functions in relative azimuthal angle is to make the function more peaked around $\Delta \phi = \pi$ 
for p+p collisions and around $\Delta \phi=0$ for Pb+Pb collisions. Note that the normalization of the  balance functions, 
Eq. \ref{eq:balnorm}, measured in a restricted part of phase-space ($p_T>2$ GeV) is modified:

\begin{eqnarray}
  \int_{-\infty}^\infty d(\Delta y) \int_0^\pi  d(\Delta \phi) B(\Delta y, \Delta \phi) < 1
  \end{eqnarray}
and similarly for the one-dimensional balance functions.

\begin{figure}[th!]
	\includegraphics[scale=0.4]{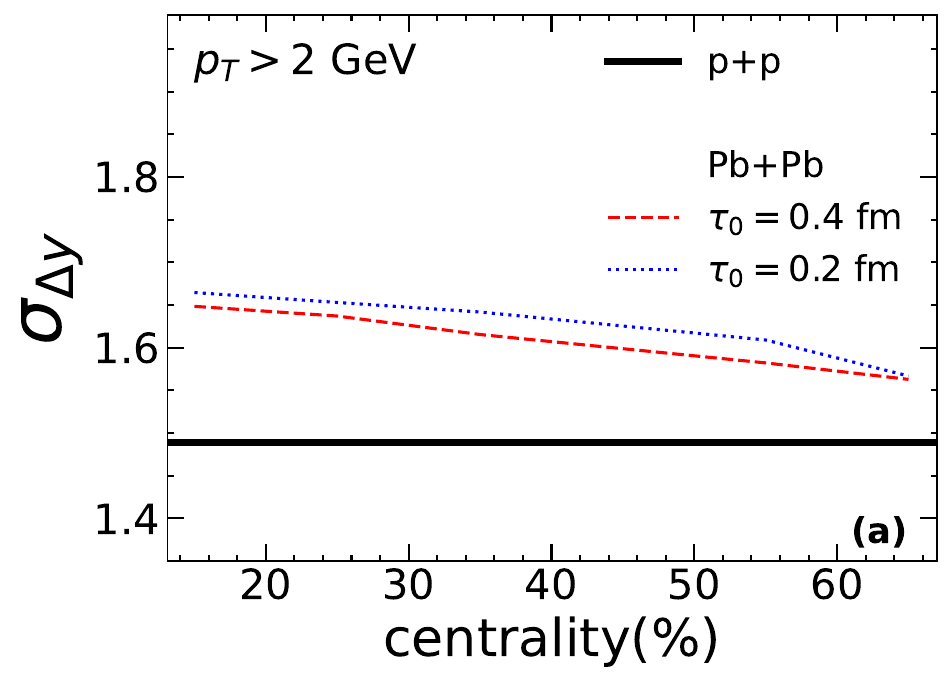} 
	\includegraphics[scale=0.4]{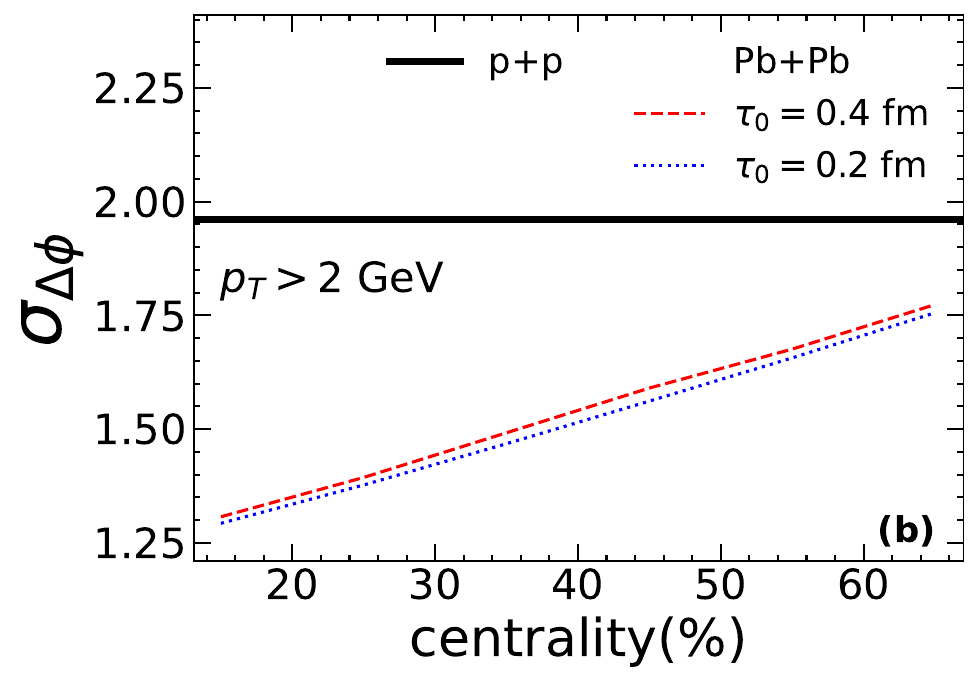}
	\caption{RMS width of the balance function in (a) relative rapidity and in (b) relative azimuthal angle 
        for $c$-$\bar{c}$ pairs produced in Pb+Pb collisions as a function 
	of collision centrality. The dotted line and dashed lines represent the results for the initial time $\tau_0=0.2$ and $0.4$fm/c respectively. 
	The horizontal solid line denotes the width of the $c-\bar{c}$ distribution in p+p collisions. All results are for quarks with $p_T>2$ GeV.}
	\label{fig:ccwy}
        \end{figure}

For the study of the balance function for charm quarks, we use the root mean square (RMS) width of the balance function in relative 
rapidity, $\sigma_{\Delta y}=\langle \Delta y^2 \rangle^{1/2}$ and not the simple width, $\langle |\Delta y| \rangle$. This makes simpler 
the discussion of different contributions to the total broadening of the balance function. The RMS width in Pb+Pb collisions is larger than 
in p+p collisions (Fig. \ref{fig:ccwy}(a)). The quark-antiquark pair undergoes additional rescattering in the fireball widening the distribution. 
The width shows a clear centrality dependence. The calculation with reduced starting time for the hydrodynamic evolution of the fireball 
($\tau_0=0.2$ fm/c instead of $0.4$ fm/c) shows a slightly larger width of the balance function. The mechanism responsible for the widening 
of the balance function in rapidity is discussed in the following section.

The width of the balance function in relative azimuthal angle, $\sigma_{\Delta \phi}= \langle \Delta \phi^2 \rangle^{1/2}$ ,
shows a strong centrality dependence (Fig. \ref{fig:ccwy}(b)). The longer the $c$-$\bar{c}$ quark pair interacts with the expanding 
fluid, the more collimated the momenta of two quarks become.

\section{Evolution of the $c$-$\bar{c}$ distribution}

\label{sec:ccevolution}

Initially, the charm quark has typically a large transverse momentum, as compared to the background thermal motion. 
During the evolution,  the quark rescatters in the medium, losing part of its energy and transverse momentum and the 
direction of the transverse momentum becomes more aligned with the transverse collective flow of the fluid. 
As a result, the distribution in relative azimuthal angle becomes peaked at $\Delta \phi = 0$ \cite{Zhu:2006er,Zhu:2007ne,Younus:2013be,Tsiledakis:2009qh,Nahrgang:2013saa,Adare:2013wca}.

\begin{figure}[th!]
	\includegraphics[scale=0.4]{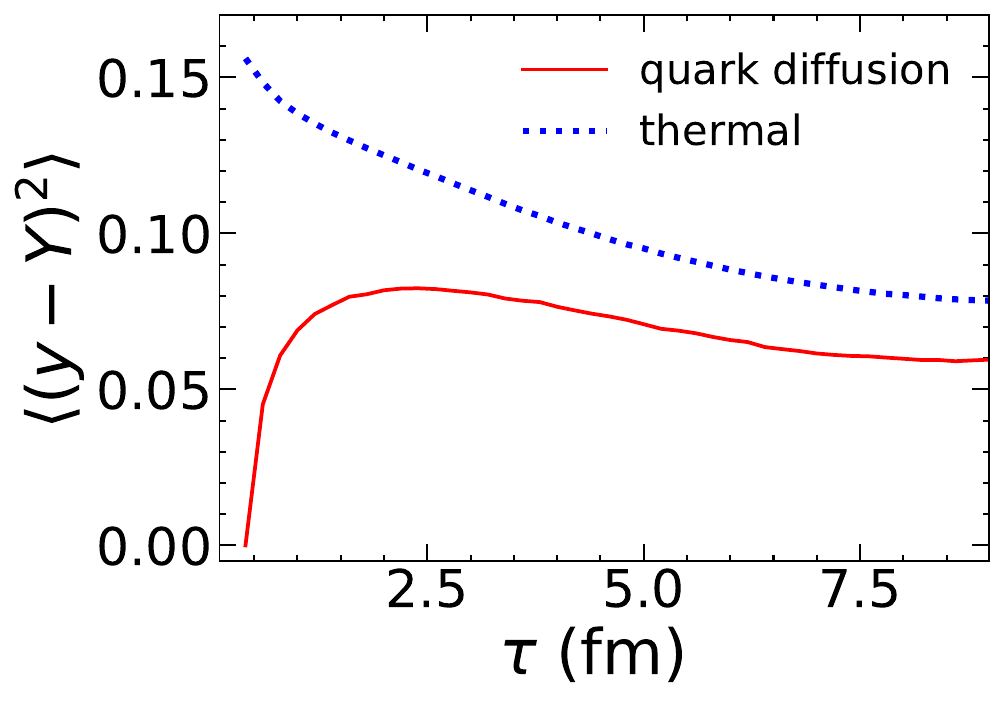} 
	\caption{The time evolution of the spread $\langle (y-Y)^2\rangle$ of the relative rapidity of the heavy quark $y$ and 
	the rapidity of the fluid for Pb+Pb collisions with $10$-$20$\% centrality (solid line). The dotted line represents the 
	thermal spreads of the relative rapidity for thermally equilibrated longitudinal velocity of the quark in the fluid 
	(Eq. \ref{eq:thermw}). }
	\label{fig:yY}
 \end{figure}

When a heavy quark is initialized in the fluid, its rapidity with respect to the fluid is zero (in the scenario when both 
the heavy quarks and the fluid follow the Bjorken scaling flow in the longitudinal direction). The rescattering with the
thermal medium leads to a spread of the relative rapidity distribution.
 The solid line in Fig.~\ref{fig:yY} illustrates the time evolution of the relative rapidity spread of  the charm quark rapidity $y$ with respect to the rapidity of the surrounding fluid $Y$, denoted as $\langle (y-Y)^2\rangle$, which is calculated from simulations of Pb+Pb collisions of 10-20\% centrality. At a specific $\tau$, the $\langle (y-Y)^2\rangle$, is computed by averaging over all the charm quarks.

Assuming a nonrelativistic thermal distribution in the longitudinal direction
\begin{equation}
  f(p_L)= \frac{1}{\sqrt{2 \pi E_T T}}\exp\left( -\frac{p_L^2}{2 E_T T }\right)\ ,
\end{equation}
where $p_L$ is the quark longitudinal momentum in the fluid rest frame and $E_T=\sqrt{m_c^2+p_T^2}$ is the transverse energy of the quark,
    the thermal distribution in the difference of the 
quark rapidity $y$ and the fluid rapidity $Y$ is:
\begin{equation}
  f_{th}(y-Y)= \sqrt{\frac{(1+\sinh(y-Y)^2)}{2 \pi E_T T}}\exp\left( -\frac{\sinh(y-Y)^2 E_T}{2 T}\right) .
  \label{eq:therm}
  \end{equation}

The average spread of the relative rapidity during the evolution of heavy quarks in the medium formed in a Pb+Pb collision 
with $10$-$20$\% centrality  is shown in Fig. \ref{fig:yY}.
The spread grows rapidly from the initial value towards the average 
thermal spread:
\begin{equation}
  \sigma_{th}^2= \langle (y-Y)^2\rangle = \int_{-\infty}^\infty d (y-Y)  (y-Y)^2 f_{th}(y-Y) \ ,
  \label{eq:thermw}
\end{equation}
shown with the dotted line in Fig. \ref{fig:yY}.
In the calculation of this thermal spreads of the relative
rapidity ($\langle (y-Y)^2 \rangle$) at a specific $\tau$, we use the average background temperature and average transverse momentum of all the considered charm quarks in the expression of the thermal distribution $f_{\text{th}}(y-Y)$ as $T$ and $p_T$ respectively.
Due to the Bjorken expansion of the system, the longitudinal momentum distribution of heavy-quarks never reaches the equilibrium 
distribution \cite{Florkowski:2013lya}. During the evolution, the average transverse momentum of the heavy quark decreases, also 
the temperature of the surrounding medium decreases. The net effect is that the  thermal spread decreases in time. The final thermal 
spread at  the freeze-out is small  and gives a very small contribution to the broadening  of the balance function in the medium.

\begin{figure}[th!]
	\includegraphics[scale=0.4]{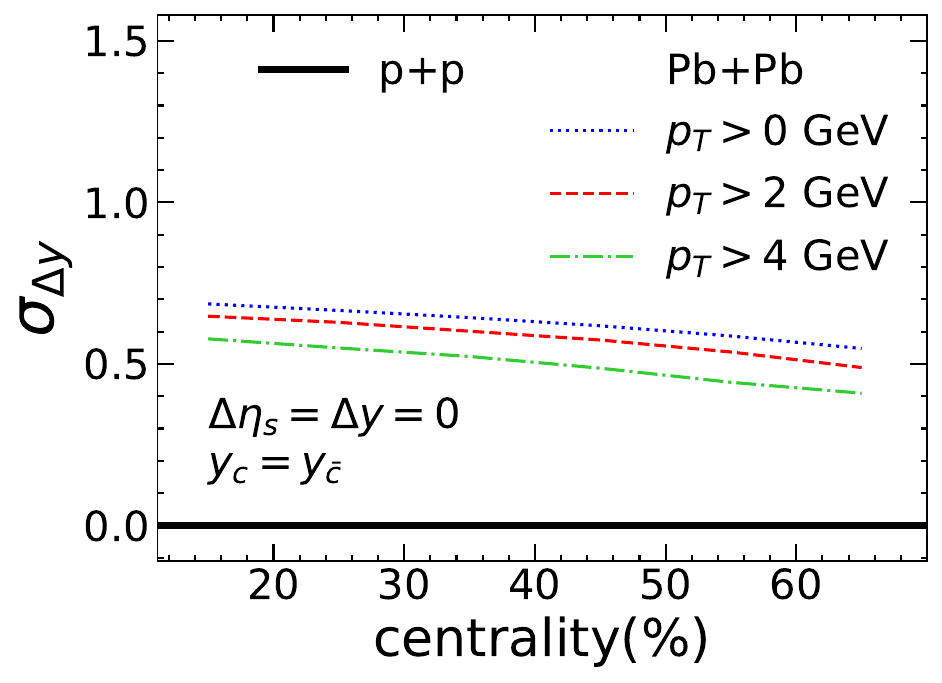} 
	\caption{The width of the charm balance function as a function of collision centrality for the initialization with both 
	quarks with the same rapidity and same position.}
	\label{fig:diffcc}
        \end{figure}

\begin{figure}[th!]
	\includegraphics[scale=0.4]{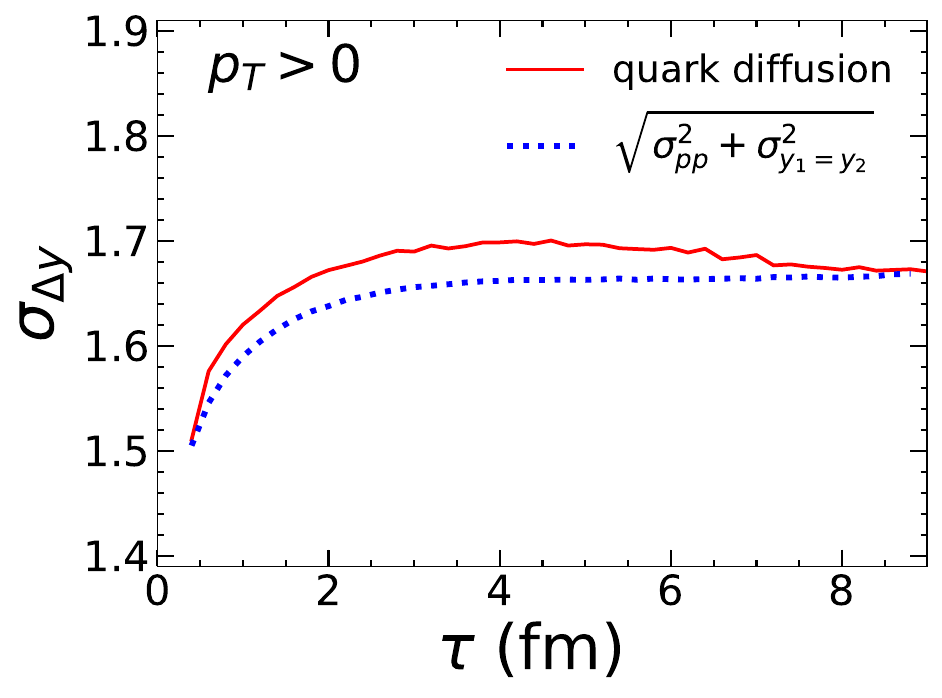} 
	\caption{The time evolution of the  width of the charm balance function as  in Pb+Pb collisions with $10$-$20$\% centrality
	 (solid line). The dotted line represents the broadening for two quarks initialized at the same rapidity.}
	\label{fig:diffcctime}
        \end{figure}

The random evolution of the quark-antiquark pair in the medium leads to a diffusion in rapidity and to an increase of the 
average rapidity separation between  the two charges \cite{Bass:2000az}. To test this effect we perform a simulation with 
both the $c$  and the  $\bar{c}$ quarks initialized at the same rapidity, $y=\frac{y_1+y_2}{2}$ and with the same 
space-time rapidity, $\eta_s=y$. Here $y_1$ and $y_2$ are the rapidities of the $c$-$\bar{c}$ pair as sampled from p+p collision 
in PYTHIA. Initially the width of the balance function is  zero  and increases in time due to the diffusion 
and thermal spread. The total broadening in Fig. \ref{fig:diffcc} is much larger than  the thermal broadening plotted Fig. \ref{fig:yY}.

The Fig.~\ref{fig:diffcctime} shows the time evolution of the width of the charm balance function in relative rapidity within a realistic simulation, as denoted by the solid line. Notice, that the width of the balance function increases strongly in the first stage of the diffusion. It shows, that the increase of 
the width of the balance function in rapidity is sensitive to the early stage of the evolution.

Additionally, the results of a simulation that considers a specific initialization scheme for charm quarks are presented ($\sigma^2_{y_1 = y_2}$) with a dotted line in Fig.~\ref{fig:diffcctime}. In this particular scheme, both the charm  and anticharm  quarks are initialized with same rapidity values ($y=\frac{y_1 + y_2}{2}$), resulting in zero initial spread in relative rapidity between charm and anticharm quarks. Since the initial distribution in relative rapidity is independent from the subsequent diffusion, the total  width is presented by incorporating the baseline width obtained from p+p collisions ($\sigma_{pp}$) through quadratic summation. 
In a boost invariant system the diffusion of a quark  or antiquark would be independent on its initial rapidity. In that case, the estimate $\sqrt{\sigma_{pp}^2+\sigma_{y_1=y_2}^2}$ of the total width  would be exact.
  It is slightly smaller than the actual 
broadening in the full calculation. In the full calculation the two quarks are initialized in two different fluid cells, with different 
rapidities. During a three-dimensional hydrodynamic evolution the fluid accelerates in the longitudinal direction, which increases 
the rapidity of the two fluid elements, leading to violation of the Bjorken scaling flow by about $3$-$5$\%. This effect increases 
the relative rapidity separation of the two fluid elements and of the two heavy quarks as well.

\section{Charm balance function for $D^0-\bar{D^0}$ mesons}

\label{sec:DD}

The charm quarks hadronize into open charm mesons and baryons.  In this work, we perform the hadronization at the freeze-out 
hypersurface of the medium which is obtained at a constant temperature of 150 MeV. We model the hadronization using the 
fragmentation model of Peterson et al. \cite{Peterson:1982ak}. Constructing the charm balance function using all open charm 
mesons could recover almost all of the balancing charm charges. In this paper we present an estimate for the balance function 
of $D^0$-$\bar{D}^0$  mesons only. The probability of producing a $D^0$ meson from a charm quark is around $40$\% \cite{ALICE:2021dhb}. 
This means that only about $40$\% of the balancing anticharm would be
recovered in $\bar{D}^0$ for each observed $D^0$, assuming full acceptance and efficiency.

The estimate of the D meson balance function reflects the dynamical evolution of the charm-anticharm quark balance function. It should be noted that the final d meson balance function can be also modified  during the hadronization process or by rescattering of the newly formed D mesons with surrounding hadrons
\cite{Ozvenchuk:2014rpa}. The sensitivity of the D meson balance function to  possible hadronization scenarios and to the hadron rescattering is important and should be studied to understand and reduce  the uncertainties in the predictions. Such
a more realistic extended investigation goes beyond this study restricted to charm diffusion in the dense phase only.

\begin{figure}[th!]
	\includegraphics[scale=0.4]{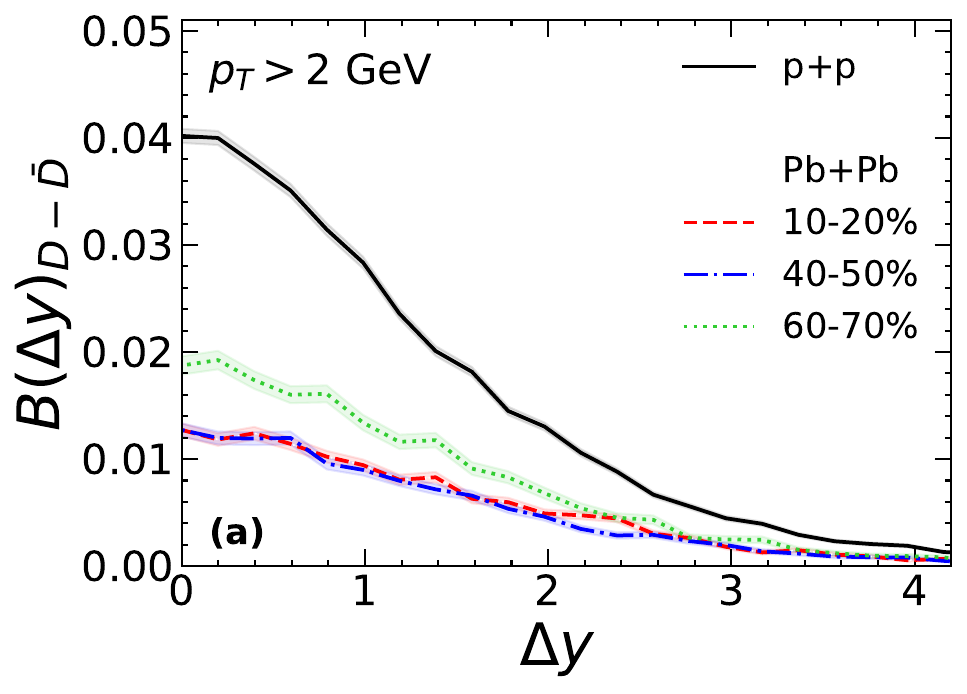}
	\includegraphics[scale=0.4]{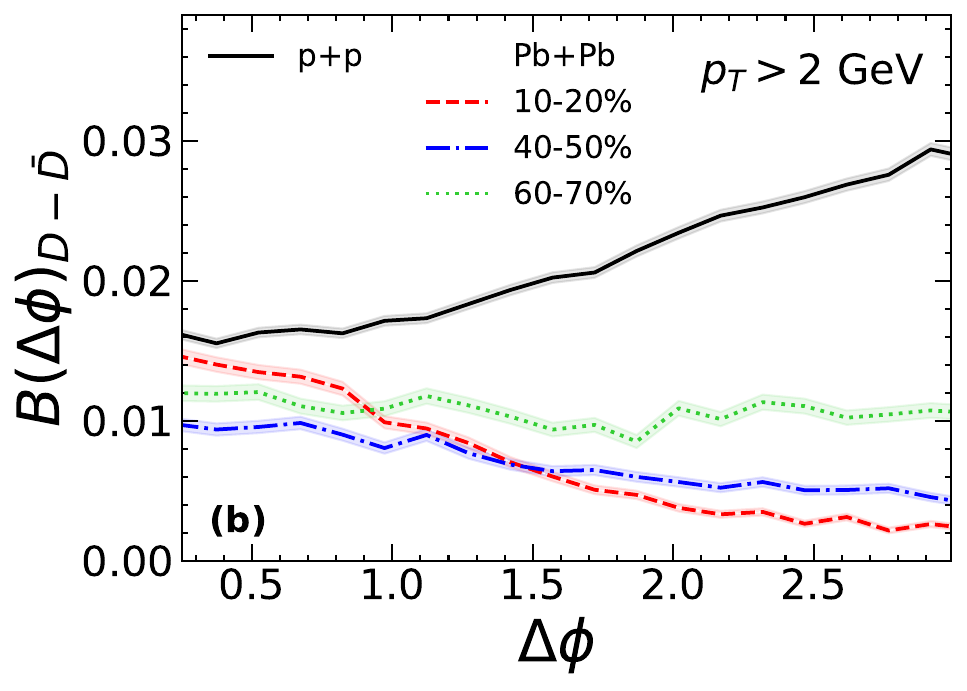}
  \caption{ The balance function in (a) relative rapidity and in (b) relative azimuthal angle for $D^0$-$\bar{D^0}$ 
    mesons with $p_T>2$GeV in Pb+Pb collisions with centrality 
  range $10$-$20$\% (dashed line), $40$-$50$\% (dashed-dotted line), and $60$-$70$\% (dotted line). The solid line represents the balance 
  function for p+p collisions.}
	\label{fig:BFyDD}
        \end{figure}
            
      
\begin{figure}[th!]
	\includegraphics[scale=0.4]{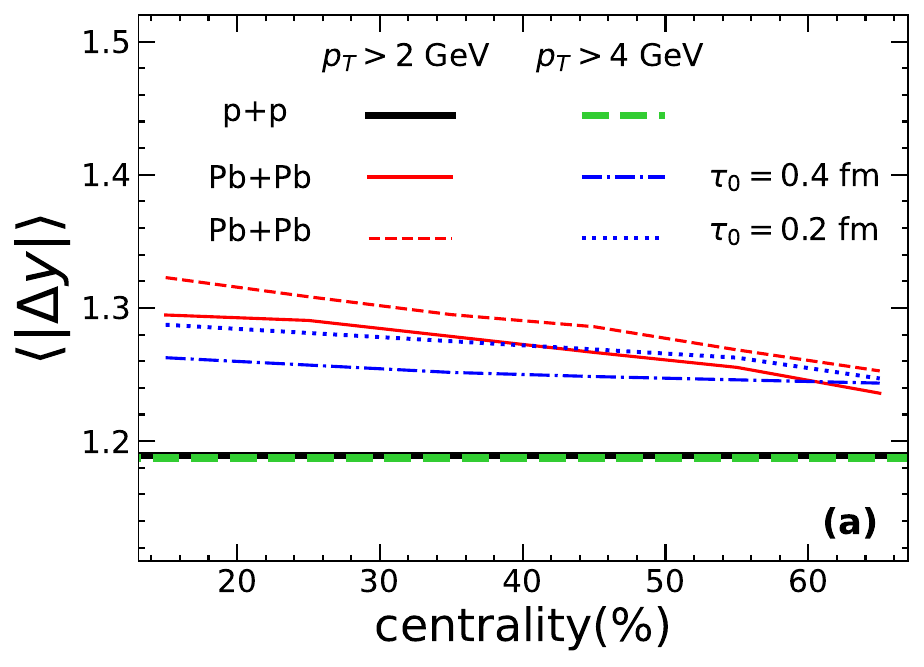}
	\includegraphics[scale=0.4]{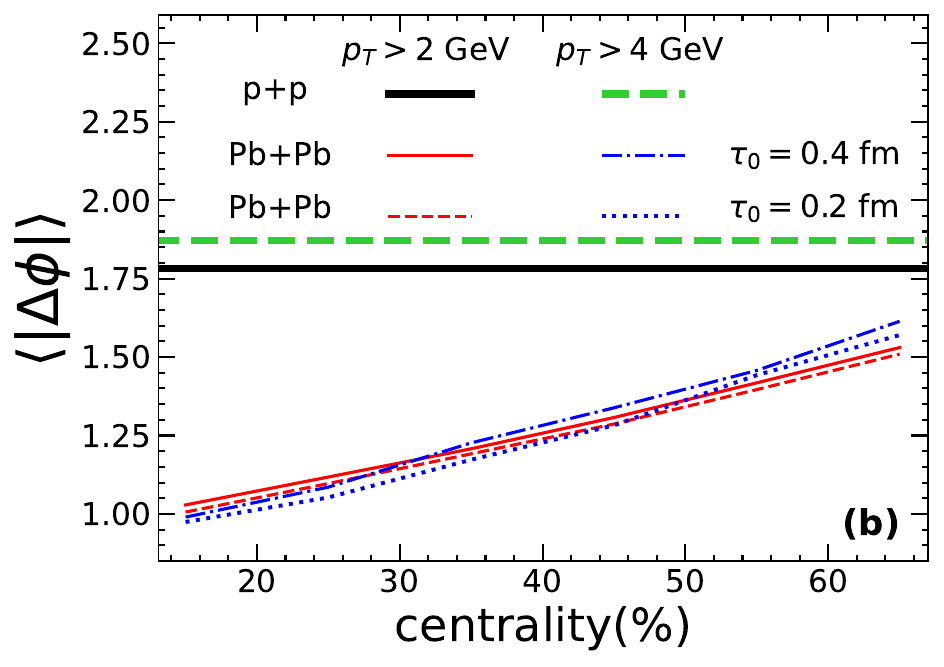} 
  \caption{ The width (a) $\langle |\Delta y |\rangle$ and (b) $\langle |\Delta \phi  |\rangle$ of the balance function 
    for $D^0$-$\bar{D^0}$ mesons as a function of collision centrality. 
    The solid and dashed line represent the results of model calculation with the initial 
    time $\tau_0=0.4$ and $\tau_0=0.2$ fm respectively for mesons of $p_T>2$ . The dashed-dotted and dotted line show the results of $p_T > 4$ GeV with the initial 
    time $\tau_0=0.4$ and $\tau_0=0.2$ fm respectively.  
    The width of the $D^0$-$\bar{D}^0$ distribution in p+p 
    collisions are denoted by solid and dashed horizontal line for meson of $p_T>2$ and $p_T>4$ GeV respectively.}
	\label{fig:widthyDD}
\end{figure}


The balance function measured for the open charm mesons $D^0$-$\bar{D}^0$  represents qualitatively the same features as the balance 
function for quarks. The balance function in relative rapidity gets broader due to the diffusion of charm quarks in the medium (Fig. \ref{fig:BFyDD}(a)).  
As expected, the normalization of the balance function is  modified, as not all the charm quarks are identified by measuring only the $D^0$ mesons. 
The balance in the relative azimuthal angle indicates that a collimation of the momenta of the $D^0$ and $\bar{D}^0$ meson occurs due to the 
interaction of the heavy quarks with the expanding medium (Fig. \ref{fig:BFyDD}(b)).

For the prediction of the  balance function for $D^0$-$\bar{D}^0$ pairs, we use the average width (as typically used by experimental groups), not 
the RMS width as for the discussion of the charm quark balance function in previous sections. The width of the balance function in relative rapidity 
increase as compared to p+p collisions (Fig. \ref{fig:widthyDD}(a)). The width, $\langle |\Delta y |\rangle$, shows some centrality dependence.  A change 
in the starting time, $\tau_0$, for the quark evolution in the quark-gluon plasma, has a sizable effect on the width of $D^0$ meson balance function. 
These effects indicate a possible sensitivity of the width of the charm meson balance function  to the dynamics of the dense matter in the collision. 
Especially interesting is the sensitivity to the rescattering of heavy quarks in the early stages of the dynamics. The balance function of $D^0$ mesons in relative 
rapidity could serve as probe of the formation time of the dense matter.

The width of the balance function in relative azimuthal angle is a very sensitive probe of the degree of thermalization of the direction of  heavy 
quarks in the medium (Fig. \ref{fig:widthyDD}(b)). The width depends strongly on the collision centrality. The collimation of  open charm mesons 
comes from the drag of the heavy quarks in the expanding medium. This mechanism is active only in the latter stages of the collision. In the early 
phase of the collision, when the transverse collective flow is not yet present, the rescattering leads to an isotropization in the relative azimuthal angle, 
but not a collimation. Therefore, the width of the balance function is not sensitive to the starting time of 
the dynamics, $\tau_0$. On the other hand, the width, $\langle \vert \Delta \phi \vert \rangle$, could serve as a clock for the total lifetime of 
the quark-gluon plasma.

\section{Effect of early dynamics  on the balance function}

\label{sec:early}

The formation time of the heavy quark pair can be estimated as $1/2 m_c\simeq 0.07$ fm/c. In the previous section are presented results assuming 
the rescattering of the heavy quarks starting at the initial time for hydrodynamics $\tau_0=0.4$ or $0.2$ fm/c. However, the evolution of the momenta 
of heavy quarks in the very early, preequilibrium  phase could change the charm balance as well \cite{Avramescu:2023qvv,Boguslavski:2023fdm}.
In the present manuscript, we estimate this effect assuming  that the dynamics of heavy quarks in the early phase is still given by the Langevin equation 
and by extrapolating the preequilibrium evolution of the density of the medium down to $\tau_f$. For the evolution of the energy density  in the earliest 
phase, $\tau<\tau_0$, we use the assumption~:
\begin{equation}
  \epsilon(\tau,x,y, \eta_s)=  \epsilon(\tau_0,x,y, \eta_s) \left(\frac{\tau_0}{\tau}\right)^\alpha \ ,
\end{equation}
with $\alpha=4/3$ (ideal hydrodynamics with conformal equation of state), $1$ (zero longitudinal pressure), or $0$ (longitudinal electric field) \cite{Vredevoogd:2008id}. 
For the  flow velocity we take the Bjorken flow, with no transverse flow for $\tau<\tau_0$.

\begin{figure}[th!]
	\includegraphics[scale=0.4]{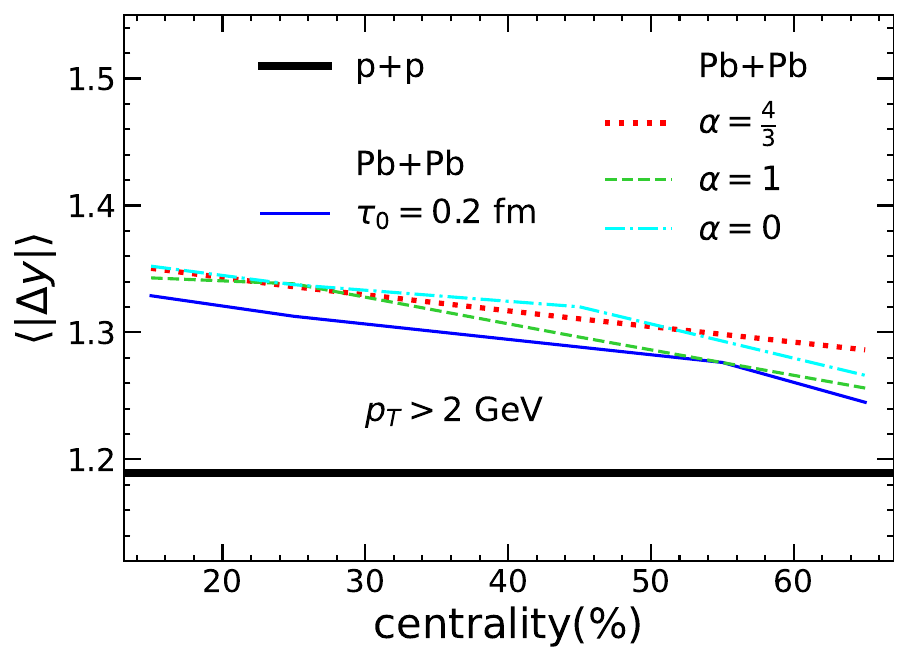} 
  \caption{ The width  $\langle |\Delta y  |\rangle$ of the balance function
    for $D^0$-$\bar{D^0}$ mesons as a function of collision centrality obtained with different assumptions on the evolution of the energy density in the early phase. 
    Results assuming a hydrodynamic evolution starting at $\tau_0=0.2$ fm/c and a  preequilibrium evolution of the energy density, 
    $\epsilon(\tau)=\epsilon(\tau_0)\left(\frac{\tau_0}{\tau}\right)^\alpha$ from $\tau_f=0.07$fm/c to $\tau_0=0.2$ fm/c are plotted for $\alpha=4/3$ (dotted line), 
    $1$ (dashed line), and $0$ (dashed-dotted line). Results obtained with a hydrodynamic evolution only, starting at $\tau_0=0.2$ fm/c, are represented with 
    the solid line.}
	\label{fig:widthyearly}
\end{figure}


We notice that the early diffusion of heavy quarks has  an effect on the width of the balance function in relative rapidity (Fig. \ref{fig:widthyearly}). 
 This observation is in agreement with the results in Sec. \ref{sec:DD}. It would be interesting to study such effect 
using specific models   of the evolution of the heavy quarks in the preequilibrium phase \cite{Sun:2019fud,Avramescu:2023qvv,Boguslavski:2023fdm} or 
including the effect of electromagnetic fields \cite{Das:2016cwd}. Further, it will be also interesting to understand how the breaking of the Bjorken scaling 
in the initialization of the charm quarks as well as a possible non-Bjorken flow velocity of the medium affect the charm balance function in relative rapidity.

\section{Summary}

\label{sec:summary}

The dynamics of charm-anticharm balancing is studied for relativistic Pb+Pb collisions. Pairs of charm and anticharm quarks are formed in the 
initial hard collisions. The two charges evolve in the dense matter of the fireball and finally hadronize into open charm mesons. The charm 
balance function allows to measure  directly the distribution of  balancing charm hadrons corresponding to the early produced charm quarks.

The charm balance function becomes broader in relative rapidity and narrower in relative azimuthal angle of the quarks. The dominant 
contribution to the broadening of the charm balance in rapidity comes from diffusion of the two quarks in rapidity. The thermal spreading of 
the quark rapidity is small. We observe a centrality dependence in the widening of the charm balance function in relative rapidity, with the largest 
broadening occurring in central collisions. The balance function in relative azimuthal angle becomes narrower after rescattering of the heavy quarks 
in the expanding medium. The collimation of the momenta of the two heavy quarks is due to transverse collective flow of the fluid. This collimation 
effect has a strong centrality dependence.

The measurements of the charm balance function in relative rapidity and in relative azimuthal angle are complementary. The narrowing of the balance 
function in relative azimuthal angle is sensitive to the rescattering of heavy quarks in latter stages of the collisions, when the transverse flow is substantial. 
The broadening of the balance function in relative rapidity is sensitive to the early phase of the dynamics, when the rescattering of the heavy quarks is the 
strongest. It would be interesting to compare the charm balance functions in rapidity and azimuthal angle obtained in model calculations to results of 
 high efficiency experiments. Such observations could constrain both the total lifetime and the formation time of the dense fireball formed in 
relativistic heavy-ion collisions.

\section*{Acknowledgment}
P.B. acknowledges support from the National Science Centre  grant  2018/29/B/ST2/00244. SC acknowledges IISER Berhampur for seed grant.
	
\bibliography{../../hydr.bib}

\end{document}